\documentclass[aps,prl,reprint]{revtex4-1}%
\usepackage{amsmath}
\usepackage{array}
\usepackage{array}
\usepackage{amsfonts}
\usepackage{amssymb}
\usepackage{graphicx}
\usepackage{color}
\usepackage{dcolumn}
\usepackage{bm}
\usepackage{hyperref}
\usepackage{epstopdf}
\begin{document}

\title{Effects of screened Coulomb interaction on spin transfer torque}
\author{Adam B. Cahaya}
\affiliation{Department of Physics, Faculty of Mathematics and Natural Sciences, Universitas Indonesia, Depok 16424, Indonesia} 
\author{Muhammad Aziz Majidi}
\affiliation{Department of Physics, Faculty of Mathematics and Natural Sciences, Universitas Indonesia, Depok 16424, Indonesia} 
\date{\today }

\begin{abstract}

In magnetic multilayer, magnetizations can be manipulated by spin transfer torque. Both spin transfer torque and its reciprocal effect, spin pumping, are governed by spin mixing conductance. The magnitude of spin mixing conductance at the interface of nearly magnetic metal has been theoretically shown to be enhanced by electron - electron interaction. However, experiments show both increasing and decreasing  values of spin mixing conductance for metals with larger electron - electron interaction. Here we take into account the effect of electron - electron interaction on the screening of the Coulomb interaction at the magnetic interface to correctly describe the experiment.

\end{abstract}

\maketitle
\section{Introduction}
Since the discovery of the giant magnetoresistance effect in magnetic multilayers, the research area of spintronics that manipulate and control spin current has emerged \cite{reviewGMR,PhysRevB.72.024426}. In a bilayer of ferromagnet insulator and non-magnetic metal, magnetizations dynamics can be manipulated by spin current and vice versa \cite{SpinCurrent}. The former phenomenon is known as spin transfer torque \cite{PhysRevB.77.224419}. On the other hand, spin pumping is interfacial spin current generation by dynamic magnetization of a ferromagnetic layer into adjacent non-magnetic metal\cite{PhysRevB.66.224403}. 
The physics of spin pumping can be understood in terms of exchange interaction between magnetization and spin-polarized conduction electron \cite{PhysRevB.68.224403}. The conduction electrons of the adjacent non-magnetic metal is spin-polarized via exchange interaction with the ferromagnetic \cite{PhysRevB.96.144434}. An adiabatic precession of the magnetization pumps a spin current from ferromagnet to non-magnetic layer with a polarization\cite{PhysRevLett.88.117601}
\begin{align}
\textbf{J}=G_{\uparrow\downarrow} \textbf{m}\times \dot{\textbf{m}} ,
\end{align}
where $\textbf{m}$ is the magnetization direction and $G_{\uparrow\downarrow}$ is a complex value with a comparably small imaginary term \cite{PhysRevB.76.104409,PhysRevLett.111.176601}.

The basic models of spin pumping employ a non-interacting description of the non-magnetic metal\cite{PhysRevB.68.224403,PhysRevB.96.144434}. While this is certainly appropriate for free-electron-like metals, it is less so for nearly magnetic metals, such as Pd and Pt. The nearly magnetic metals are characterized by large Stoner enhancement 
\begin{align}
SE=\frac{1}{1-U\mathcal{N}(\epsilon_F)}
\end{align}
in their magnetic susceptibilities \cite{PhysRevB.50.7255}. Here $U$ is Hubbard parameter that represent the electron-electron interaction strength and $\mathcal{N}(\epsilon_F)$ is the density of state at Fermi energy. The effects of large Stoner enhancement on magnetic susceptibility have been thoroughly studied \cite{PhysRevB.50.7255,Zellermann_2004,Povzner2010}. However, the studies exploring the effects of electron-electron interaction on spin mixing conductance are still few \cite{PhysRevB.88.054423}. Ref.~\onlinecite{PhysRevB.67.144418} predicts that spin mixing conductance is proportional to the square of Stoner enhancement 
\begin{align}
G_{\uparrow\downarrow}\propto SE^2. \label{Eq.3}
\end{align}
However, this theoretical prediction does not fit quantitatively well with Ref.~\onlinecite{PhysRevLett.112.197201}. Furthermore Ref.~\onlinecite{PhysRevB.94.014414} shows that the spin pumping into Pd generates smaller spin current than into Pt even though the Stoner enhancement of Pd is larger. 

The spin mixing conductance also governs the reciprocal effect, the spin transfer torque. When the metallic layer has a finite spin accumulation $\boldsymbol{\mu}$, which represent the difference of spin dependent electro-chemical potential, there is a spin current transfer from the non-magnetic interface into the ferromagnetic interface, with polarization that can be written in term of spin mixing conductance \cite{PhysRevB.77.224419}. The generated spin-transfer torque is 
\begin{align}
\boldsymbol{\tau}= G_{\uparrow\downarrow}\textbf{m}\times\left(\textbf{m}\times \boldsymbol{\mu}\right) \label{Eq.stt}.
\end{align}
 
In equilibrium, the spin currents associated spin transfer torque balances the spin pumping. In spin Seebeck effect, the balance is destroyed by a thermal gradient \cite{PhysRevB.81.214418}. The net spin current is then converted into electromotive force by the spin-orbit interaction of the non-magnetic layer. A spin Seebeck device require a nearly magnetic metal, such as Pd and Pt, as a non-magnetic layer that convert the spin current into electric current \cite{7111309}. Therefore, a better understanding of spin mixing conductance of nearly magnetic metal is required. 

In this article, we analyze the effect of the screened-Coulomb interaction on the spin transfer torque. We first analyze the screening of the exchange interaction on nearly magnetic metal. We then validate the expression of spin transfer torque that arise from the exchange interaction between the magnetic moment of ferromagnetic layer and the spin of conduction electron in nearly magnetic metal that has a finite spin accumulation. Finally, we show the effect of the screened-Coulomb interaction on the spin mixing conductance.

\begin{figure}[h]\includegraphics[width=\columnwidth]{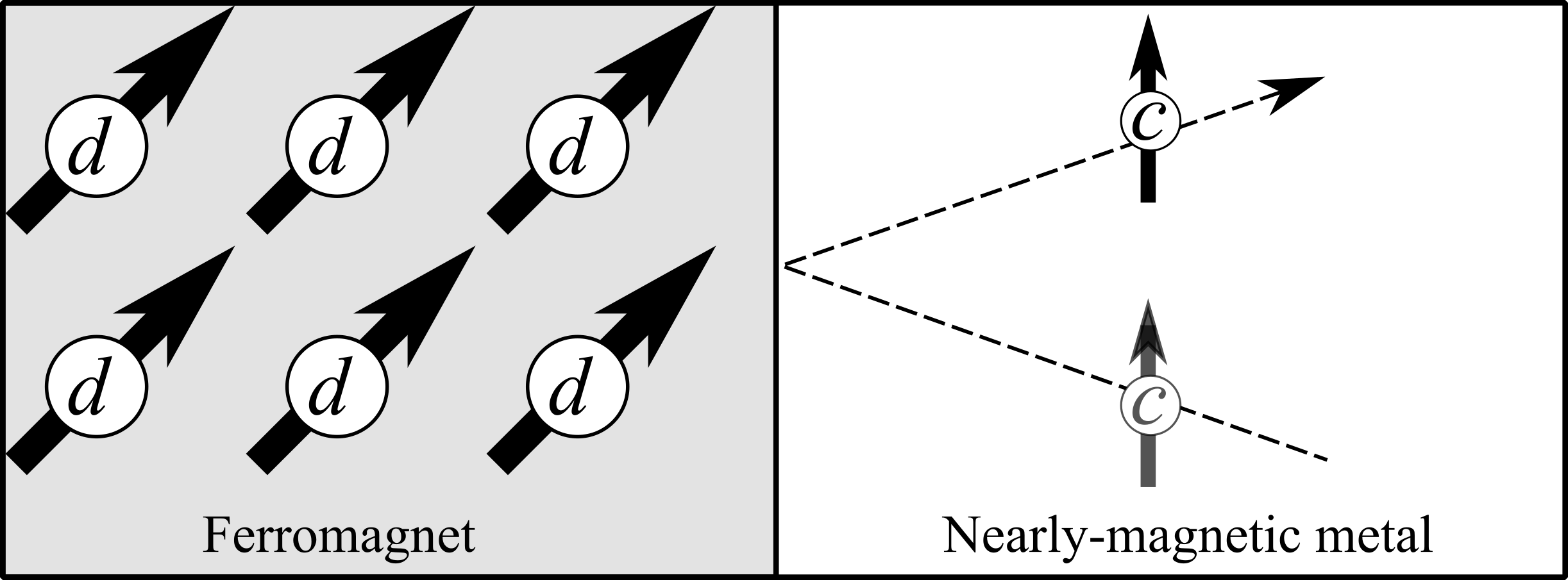}\caption{The interface of magnetic and nearly magnetic metal can be modeled as a ferromagnetic layer that consist of magnetic ions and an interacting conduction electron with spin accumulation. }\label{Fig.model}\end{figure}

\section{Screened-exchange interaction with electron-electron interaction correction}
In second quantization, the interactions in nearly magnetic system near the interface as illustrated in Fig.~\ref{Fig.model} can be written with the following  $s-d$ Hamiltonian with Hubbard interaction
\begin{align}
&H=\sum_{\textbf{p}\alpha} \epsilon_\textbf{p}a_{\textbf{p}\alpha}^\dagger a_{\textbf{p}\alpha}+U\sum_{\textbf{pqk}} a_{\textbf{p}+\textbf{q}\uparrow}^\dagger a_{\textbf{p}\uparrow} a_{\textbf{k}-\textbf{q}\downarrow}^\dagger a_{\textbf{k}\downarrow} \notag\\
&-\frac{\hbar}{2}\sum_{\textbf{p}\alpha\beta} \boldsymbol{\mu}\cdot\boldsymbol{\sigma}_{\alpha\beta} a_{\textbf{p}\alpha}^\dagger a_{\textbf{p}\beta}\notag\\
&- J_U \sum_{\textbf{pq}\alpha\beta} \sum_{j\alpha'\beta'}d_{j\alpha'}^\dagger d_{j\beta'}\boldsymbol{\sigma}_{\alpha'\beta'} \cdot\boldsymbol{\sigma}_{\alpha\beta} a_{\textbf{p}+\textbf{q}\alpha}^\dagger a_{\textbf{p}\beta},\label{Eq.Hamiltonian}
\end{align}
where  $d_{j\alpha}^\dagger (d_{j\alpha})$ is the creation (annihilation) operator of $d$-electron with spin $\alpha$, $a_{\textbf{p}\alpha}^\dagger (a_{\textbf{p}\alpha})$ is the creation (annihilation) operator of conduction electron with wave vector $\textbf{p}$ and spin $\alpha$,  $\boldsymbol{\sigma}$ is Pauli vectors, $\epsilon_\textbf{p}=\hbar^2p^2/2m$ is the energy of conduction electron. The second term is the electron-electron interaction, characterized by the Hubbard parameter $U$. 
The third term is the spin-dependent energy shift due to spin accumulation $\boldsymbol{\mu}$. 
In magnetic multilayer, there is a spin accumulation on the nearly magnetic metal that accommodates non-local spin transfer \cite{PhysRevApplied.8.064023,doi:10.1088/1468-6996/9/1/014105}.
The last term is the \textit{s-d} exchange interaction of conduction electron with localized spin \[\sum_{j\alpha'\beta'} d_{j\alpha'}^\dagger d_{j\beta'}\boldsymbol{\sigma}_{\alpha'\beta'}\equiv \textbf{S}\] with exchange constant $J_U$
\begin{align}
J_U=&\int d\textbf{r}_1d\textbf{r}_2 \psi^*_{\textbf{k}_F}(\textbf{r}_1)\psi^*_{d}(\textbf{r}_2) 
V\left(\left|\textbf{r}_1-\textbf{r}_2\right|\right)
\psi_{\textbf{k}_F}(\textbf{r}_2)\psi_{d}(\textbf{r}_1)  \label{Eq.exchange},
\end{align}
where $\psi^*_{d}(\textbf{r})=\sum_{lm}R(r)Y_{lm}{\hat{r}}$ is the wave function of the $d$-electron and $R(r)=\sqrt{8\zeta^7/45}r^2e^{-\zeta r}$ is the Slater wave function. 

While Ref.~\onlinecite{PhysRevB.67.144418} has discussed the effect of the electron-electron interaction on spin mixing conductance, the effect on the exchange constant $J_U$ was overlooked. We need to take it into account the $U$-dependency of $J_U$ to give a more accurate estimation.
The dependency of $J_U$ to $U$ arises from the screening of Coulomb interaction. A screened Coulomb interaction can be expressed in term of Yukawa potential \begin{equation}
V(\textbf{r})=\frac{e^2\exp\left(-\lambda r\right)}{4\pi\varepsilon_0 r},
\end{equation}
where screening constant $\lambda$ is 
\begin{align}
\lambda^2=& q^2V(\textbf{q})\frac{\Gamma(\textbf{q})}{1-U\Gamma(\textbf{q})},\\
\Gamma(\textbf{q})=&\lim_{q\ll k}
\sum_{\textbf{k},\boldsymbol{\sigma}}\frac{f_{\textbf{k},\boldsymbol{\sigma}}-f_{\textbf{k}+\textbf{q},\boldsymbol{\sigma}}}{E_{\textbf{k}+\textbf{q}}-E_{\textbf{k}}+i0^+},
\end{align}
and $V(\textbf{q})$ is the Fourier transformation of $V(\textbf{r})$. $\lambda$ is related to density of state at Fermi energy $\mathcal{N}(\epsilon_F)$ of non-interacting metal as \citep{BookKim}
\begin{align}
\lambda^2=&\frac{e^2\varepsilon_0^{-1}\mathcal{N}(\epsilon_F)}{1-U\mathcal{N}(\epsilon_F)} .
\end{align}
Here we note that $\lambda(U=0)\equiv\lambda_0\sqrt{e^2\epsilon_0^{-1}\mathcal{N}(\epsilon_F)}$. Substituting the spherical harmonic expansion of screened Coulomb potential \cite{Racah,JIAO2015140,Bagci2018}
\begin{align}
\frac{e^{-\lambda \left|\vec{r}_1-\vec{r}_2 \right|}}{\left|\vec{r}_1-\vec{r}_2 \right|} =&
\lambda \sum_{l=0}^{\infty} i_l\left(\lambda r_<\right)k_l\left(\lambda r_>\right) 
\sum_{m=-l}^{l} Y_{lm}\left(\boldsymbol\Omega_1 \right) Y^*_{lm}\left(\boldsymbol\Omega_2 \right) , \label{Eq.TaylorYukawa}
\end{align}
into Eq.~\ref{Eq.exchange}, we arrive at the expression for $J_U$

\begin{align}
J_U=\frac{e^2\lambda^2}{\varepsilon_0} \int_0^\infty r^2dr \int_{0}^{\infty} r'^2dr'  R(r) i_2\left(\lambda r_>\right)     R(r') k_2\left(\lambda r_<\right).
\end{align}
Here $r_>=\mathrm{max}(r,r')$, $r_<=\mathrm{min}(r,r')$. $i_n$ and $k_n$ are the modified Spherical Bessel functions of the first and second kind, respectively. 
For a well-localized spin ($\zeta\gg k_F$) the value  of $J_U$ can be shown to be as follows
\begin{align}
J_U= \frac{e^2}{\varepsilon_0\lambda^2} \frac{1+ 8\frac{\zeta}{\lambda} + 27\frac{\zeta^2}{\lambda^2} + 48\frac{\zeta^3}{\lambda^3} + \frac{219\zeta^4}{5\lambda^4} + \frac{72\zeta^5}{\lambda^5} + \frac{\zeta^6}{5\lambda^6}}{\left(1+\frac{\zeta}{\lambda}\right)^8}. \label{JU}
\end{align}

\begin{figure} \centering \includegraphics[width=\columnwidth]{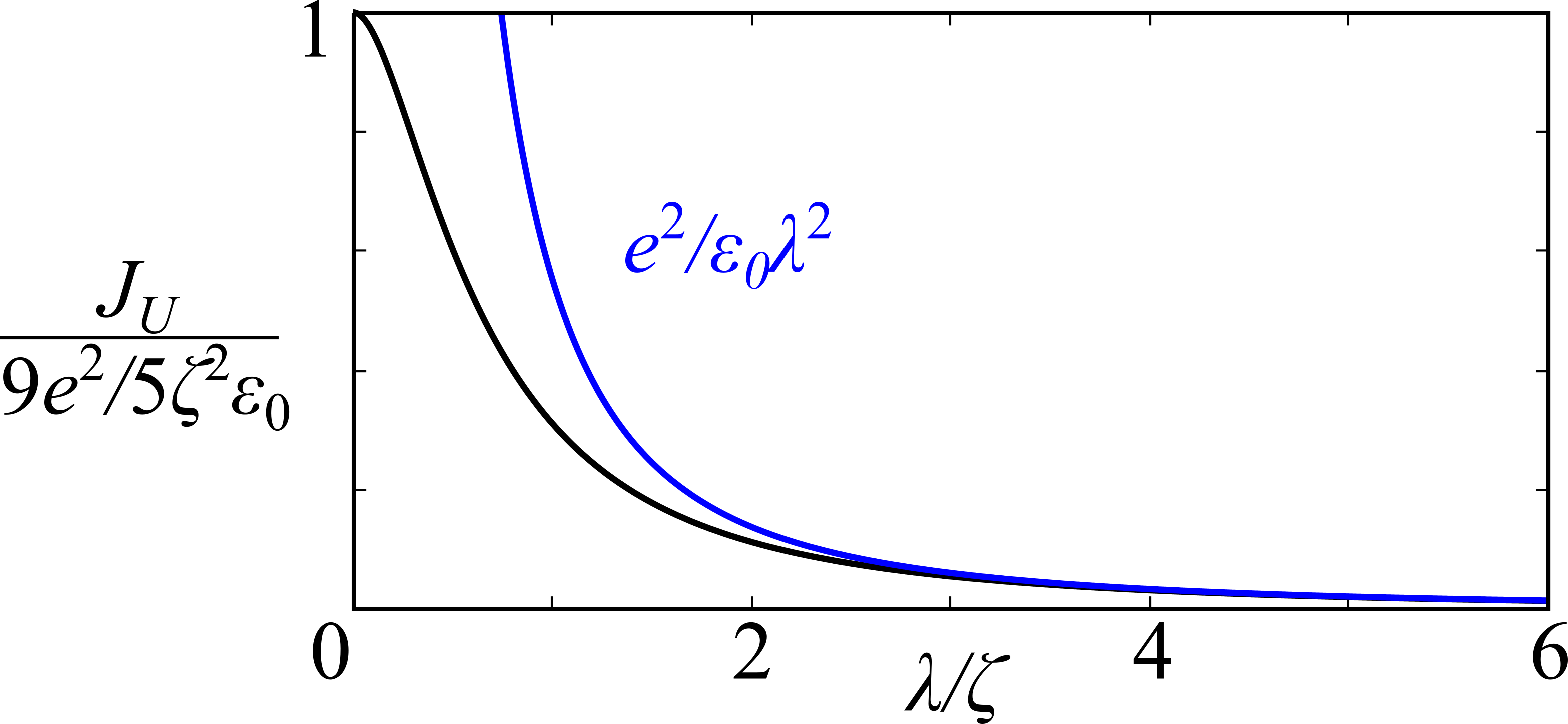}
\caption{[Color online] Exchange constant $J_U$ as a function of $\lambda/ \zeta$ (black line). For large $\lambda$, the exchange constant approachs $J_U=\frac{e^2}{\varepsilon_0\lambda^2} =J_0\left(1-U\mathcal{N}(\epsilon_F)\right) $ (blue line)}\label{Fig.exchange} \end{figure}

When the ferromagnetic layer is metallic, such as Permalloy (Py)~\cite{PhysRevB.94.014414} we can take a strong screening limit ($\lambda\gg \zeta$). In this case, the value of $J_U$ approaches the following value as illustrated in Fig.~\ref{Fig.exchange}.

\begin{align}
\lim_{\lambda\gg \zeta}J_U= \frac{e^2}{\varepsilon_0\lambda^2} =J_0\left(1-U\mathcal{N}(\epsilon_F)\right), \label{JUstrong}
\end{align}
where $J_0=(\mathcal{N}(\epsilon_F))^{-1}$ is the exchange constant for $U=0$. Here, one can see that large Stoner enhancement greatly reduce the exchange interaction. On the other hand, when the ferromagnetic layer is insulator, such as Y$_3$Fe$_5$O$_{12}$ (YIG)~\cite{PhysRevLett.112.197201}, the screening is weaker \cite{PhysRevB.90.165105,Racah}.

\section{Spin-accumulation-induced anisotropic spin-density}
In linear response regime, the exchange interaction dictates that the spin density $\boldsymbol{\sigma}$ of the conduction electron respond linearly to perturbation due to exchange interaction
\begin{align}
{\sigma}_i(\textbf{r})=J_U \int d\textbf{r} dt \chi_{ij} (\textbf{r}-\textbf{r}',t-t') S_{j}(\textbf{r}',t'),
\end{align}
where $i,j\in\left\{x,y,z\right\}$. The susceptibility
\begin{align}
\chi_{ij} (\textbf{r},t)=& \frac{i}{\hbar}\Theta (t) \left< \left[\sigma_i(\textbf{r},t),\sigma_j(\textbf{0},0)\right]\right>
\end{align}
can be determined by evaluating its time derivation
\begin{align}
\frac{\partial\chi_{ij}(\textbf{r},t)}{\partial t}=\frac{i}{\hbar}\Theta (t) \left< \left[\frac{1}{i\hbar}\left[\sigma_i(\textbf{r},t),H\right],	\sigma_j(\textbf{0},0)\right]\right>.
\end{align}
By substituting the Hamiltonian in Eq.~\ref{Eq.Hamiltonian} and writing the susceptibility as
$
\chi_{ij}(\textbf{r},t)= \sum_{\textbf{pq}} e^{i\textbf{q}\cdot \textbf{r}-i\omega t}\chi_{ij}
(\textbf{p},\textbf{q},\omega),
$
we can derive the exact expression of $\chi_{ij}$ in the static limit $\omega\to0$ for all $i,j$ combination

\begin{align}
&\sum_\textbf{p}\left(
\begin{array}{ccc}
\chi_{xx}(\textbf{p},\textbf{q},0) & \chi_{xy}(\textbf{p},\textbf{q},0) & \chi_{xz}(\textbf{p},\textbf{q},0) \\
\chi_{yx}(\textbf{p},\textbf{q},0) & \chi_{yy}(\textbf{p},\textbf{q},0) & \chi_{yz}(\textbf{p},\textbf{q},0) \\
\chi_{zx}(\textbf{p},\textbf{q},0) & \chi_{zy}(\textbf{p},\textbf{q},0) & \chi_{zz}(\textbf{p},\textbf{q},0)
\end{array}
\right)\notag\\
&= 
\left(
\begin{array}{ccc}
\chi_1(q) & \hbar\mu_z \chi_2(q) & \hbar\mu_y \chi_2(q) \\
-\hbar\mu_1 \chi_2(q) & \chi_1(q) & \hbar\mu_x \chi_2(q)\\
\hbar\mu_y \chi_2(q) & -\hbar\mu_x \chi_2(q) & \chi_1(q)
\end{array}
\right)
\end{align}
such that
\begin{align}
\chi_{ij}(\textbf{r},t)=&\sum_\textbf{q} e^{i\textbf{q}\cdot \textbf{r}-i\omega t}\sum_{\textbf{p}}\chi_{ij}(\textbf{p},\textbf{q},0) \notag\\
=& \delta (t)\sum_\textbf{q} e^{i\textbf{q}\cdot \textbf{r}}\Big(\delta_{ij}\chi_1(q) +\hbar\epsilon_{ijk}\mu_k \chi_2(q)\Big).
\end{align}

One can see that the susceptibility is anisotropic. In the limit of small spin-accumulation $\hbar\mu\ll \epsilon_F$, the induced spin density takes the following form 
\begin{align}
\boldsymbol{\sigma}(\textbf{r})=\sum_{\textbf{k}} e^{i\textbf{k}\cdot \textbf{r}}J_U\left(\chi_1(k) \textbf{S}  + \hbar \chi_2(k) \textbf{S}\times \boldsymbol{\mu} \right), \label{Eq.20}
\end{align}
where $\chi_1(k)$, $\chi_2(k)$ and their inverse Fourier transform $\chi_1(r)$, $\chi_2(r)$ are
\begin{align}
\chi_1(k)=&\frac{\chi_0(\textbf{k})}{1-U\chi_0(\textbf{k})},\notag\\
\chi_2(k)=&\frac{\phi_0(\textbf{k})}{\left(1-U\chi_0(\textbf{k})\right)^2}\notag,\\
\chi_{1,2}(r)=& \int \frac{d\textbf{k}}{(2\pi)^3}e^{i\textbf{k}\cdot\textbf{r}} \chi_{1,2}(k) \label{Eq.chi12} .
\end{align}
Here $\chi_0$ is the static susceptibility of a metal with $U=0$ 
\begin{align}
\chi_0(\textbf{k})=& \lim_{\eta\to 0}\sum_{\textbf{p}}\frac{f_{\textbf{p}}-f_{\textbf{p}+\textbf{k}}}{\epsilon_{\textbf{p}+\textbf{k}}-\epsilon_{\textbf{p}}+i\eta}\notag\\
=& \mathcal{N}(\epsilon_F) \left(1+\frac{4k_F^2-q^2}{4k_Fk}\log\left|\frac{k+2k_F}{k-2k_F}\right|\right)
\end{align}
and
\begin{align}
\phi_0(k)=&\lim_{\eta\to 0}\sum_{\textbf{p}}\frac{f_{\textbf{p}}-f_{\textbf{p}+\textbf{k}}}{\left(\epsilon_{\textbf{p}+\textbf{k}}-\epsilon_{\textbf{p}}+i\eta\right)^2}\notag\\
=&\mathcal{N}^2(\epsilon_F)\frac{\pi^2k_F^2}{\hbar } \frac{\Theta(2k_F-k)}{k}. \label{Eq.23}
\end{align}

Incidentally, $\chi_0$ and $\phi_0$ can also be obtained by taking the limit of small $\hbar\omega \ll \epsilon_F$ to the dynamic susceptibility of metal with non-interacting conduction electrons $\lim_{\hbar\omega \ll \epsilon_F}\chi(\textbf{k},\omega)=\chi_0(k)+i\omega \phi_0(k) $.

%\begin{align} \lim_{\hbar\omega \ll \epsilon_F}\chi(\textbf{k})=& \lim_{\hbar\omega \ll \epsilon_F}\lim_{\eta\to 0}\sum_{\textbf{p}}\frac{f_{\textbf{p}}-f_{\textbf{p}+\textbf{k}}}{\epsilon_{\textbf{p}+\textbf{k}}-\epsilon_{\textbf{p}}+\hbar\omega+i\eta}\notag\\=& \chi_0(k)+i\omega \phi_0(k) \end{align}

\section{Spin transfer torque by spin accumulation}

\begin{figure}[h]
\includegraphics[width=\columnwidth]{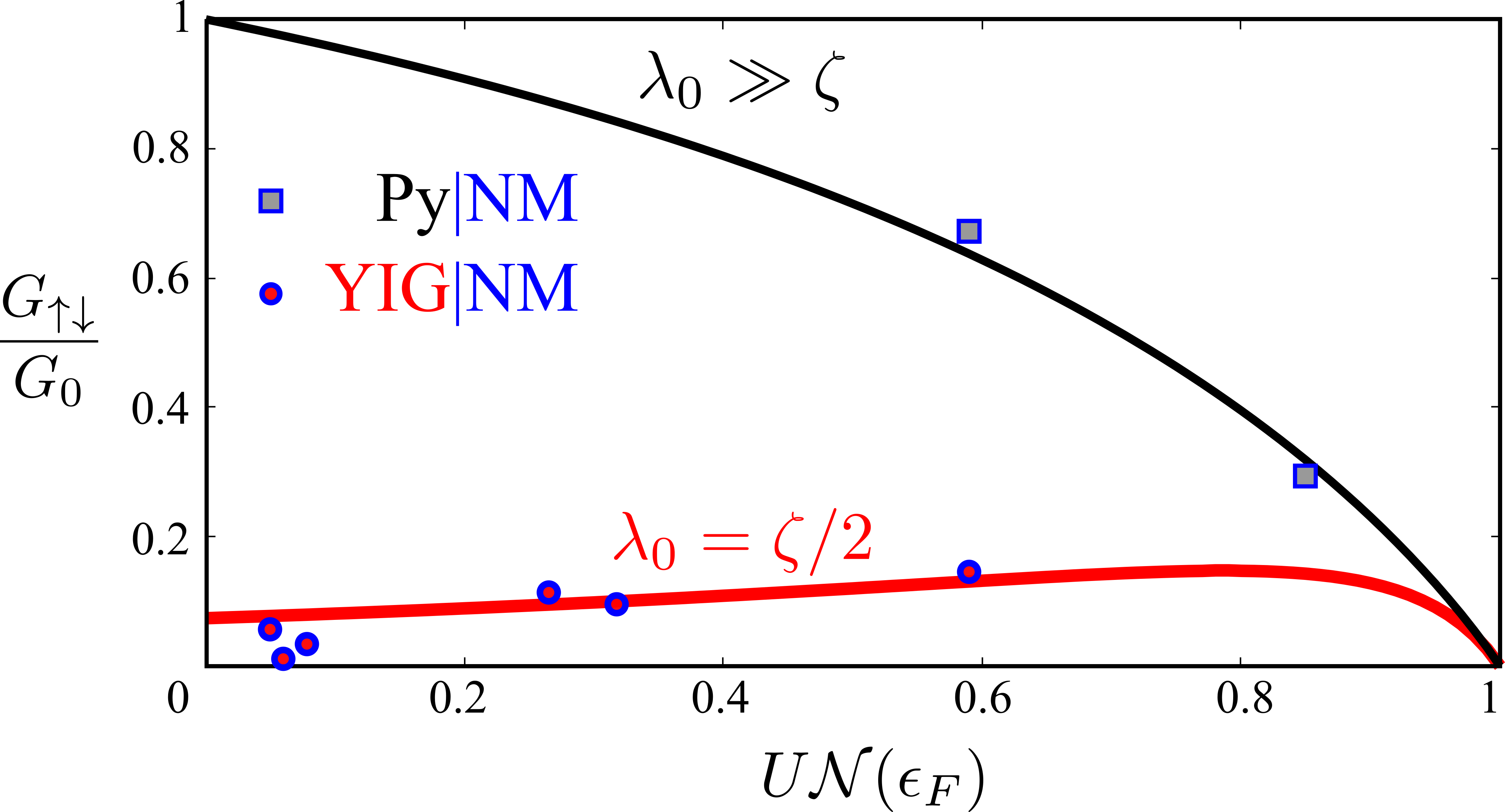}
\caption{[Color online] Enhancement of spin mixing conductance $G_{\uparrow\downarrow}$ as a function of $U\mathcal{N}(\epsilon_F)$. $G_{0}=0.5$ \AA$^{-2}$ For nearly magnetic metals with $U\mathcal{N}(\epsilon_F)\to 1$, the spin mixing conductance reduces to zero. We can see that data of Ref~\onlinecite{PhysRevB.94.014414} and \onlinecite{PhysRevLett.112.197201} are for $\lambda_0\gg\zeta$ (black line) and $\lambda_0=\zeta/2$ (red line), respectively. Ref~\onlinecite{PhysRevB.94.014414} used Permalloy (Py) as the ferromagnetic layer and Py and Pd as nearly magnetic metal layer (NM). On the other hand, Ref~\onlinecite{PhysRevLett.112.197201} used Y$_3$Fe$_5$O$_{12}$ (YIG) as the ferromagnetic layer and Au, Ag, Cu, Ta, W and Pt as nearly magnetic metal layer (NM). The values of $U\mathcal{N}(\epsilon_F)$ for NM are taken from Ref.~\onlinecite{PhysRevB.50.7255}}
\label{Fig.U}
\end{figure}

Torque acting on magnetic moment $\textbf{M}=-M_s\textbf{S}/|S|$ can be obtained as 
\begin{align}
\tau= \frac{1}{i\hbar} \left[\textbf{M},-J_U\textbf{S}\cdot \boldsymbol{\sigma} (\textbf{r}=0)\right]. \label{Eq,24}
\end{align}
by substituting Eq.~\ref{Eq.20} into Eq.~\ref{Eq,24} we arrive at the following spin-transfer torque that has a similar form with Eq.~\ref{Eq.stt}.
\begin{align}
\boldsymbol{\tau}=& \textbf{S}\times \left(\textbf{S}\times\boldsymbol{\mu}\right) J_U M_s\chi_2(\textbf{r}=0)\notag\\
=& \textbf{S}\times \left(\textbf{S}\times\boldsymbol{\mu}\right) J_U M_s\int \frac{d\textbf{k}}{(2\pi)^3}\frac{\phi_0(k)}{\left(1-U\chi_0(k)\right)^2}\label{Eq.25}
\end{align}
The spin-transfer torque can be obtained by substituting Eq.~\ref{Eq.23} into Eq.~\ref{Eq.25}.
The enhancement of spin mixing conductance can be seen from the enhancement of spin-transfer torque $\boldsymbol{\tau}$.
\begin{align}
\frac{G_{\uparrow\downarrow}}{ G_{0}}=\frac{\tau}{\tau_{U=0}}= \frac{J_U}{J_0} \int_0^{2k_F} \frac{kdk}{2k_F^2}\left(\frac{1}{1-U\chi_0(k)}\right)^2. \label{Eq.spinmixing}
\end{align}
where the spin mixing conductance $G_0$ for non-interacting electron gas is well studied in Refs.~\onlinecite{PhysRevB.96.144434,PhysRevB.67.144418}. The relative magnitude of $G_{\uparrow\downarrow}$ is shown in Fig.~\ref{Fig.U}.

Here we note that Eq.~\ref{Eq.3} arises if the effect of exchange interaction is overlooked and the integral in Eq.~\ref{Eq.spinmixing} is oversimplified.
\[\frac{G_{\uparrow\downarrow}}{ G_{0}}\sim \frac{J_U}{J_0} \frac{1}{\left(1-UN(\epsilon_F)\right)^2}\]
In this case the Ref.~\onlinecite{PhysRevLett.112.197201} and Ref.~\onlinecite{PhysRevB.94.014414} seems to yield contradictive result.  By taking into account the changes of $J_U$ and numerically evaluating Eq.~\ref{Eq.spinmixing}, we  show that the discrepancy  can be explained in term of different strength of screening at the interface.

In the limit of $U\to0$, the spin mixing conductance is determined by $\chi_2=\phi_0$. In this case, our result approaches the value of spin mixing conductance that is theoretically derived for spin pumping in Ref.~\onlinecite{PhysRevB.96.144434}. This convergence is a proof of reciprocal relation between spin-transfer torque and spin pumping.
We note here that if $J_U$ is independent of $U$, the spin mixing conductance will monotonically increase as a function of the Stoner enhancement factor as predicted by  Ref.~\onlinecite{PhysRevB.67.144418}. However, the dependency of $J_U$ to $U$ (see Eq.~\ref{JU}) suppressed the spin mixing conductance as shown in Fig.~\ref{Fig.U}. 

Fig.~\ref{Fig.U} show the spin mixing conductance for the interface of ferromagnet and nearly magnetic metal with various values of $U\mathcal{N}(\epsilon_F)$. The red line shows the quantitative agreement of our result and the spin mixing conductance of a bilayer of a insulating ferromagnet (Y$_3$Fe$_5$O$_{12}$ (YIG)) and a nearly magnetic metal (Au, Ag, Cu, Ta, W or Pt). On the other hand, black line shows the spin mixing conductance of a bilayer of a metallic ferromagnet (Permalloy (Py)) and nearly magnetic metal (Pd or Pt)\cite{PhysRevB.94.014414}. We can see that data of Py | NM match strong-screening case with $\lambda_0\gg\zeta$, because the interface is metallic. On the other hand, YIG | NM matches weak-screening case with $\lambda_0=\zeta/2$.

\section{Conclusion}
To summarize, we discuss the effect of screened-Coulomb interaction on the spin transfer torque at the interface of the ferromagnet and the nearly magnetic metal. As the metal becomes nearly magnetic, the electron-electron interaction, characterized by the Hubbard parameter $U$, and the screening of the Coulomb interaction increase.
To correctly describe $G_{\uparrow\downarrow}$, we take into account the $U$-dependency of exchange constant in Eq.~\ref{JU} that arise from the screening of Coulomb interaction at the interface. 
The large electron-electron interaction of nearly magnetic metals affect the spin mixing conductance through the exchange constant $J_U$ and the spin susceptibilities $\chi_1,\chi_2$ (Eq,~\ref{Eq.chi12}).

We show the reciprocal relation between spin-transfer torque and spin pumping in the small spin accumulation limit and showed that $\chi_2$, the susceptibility  that corresponds to the spin mixing conductance for spin-transfer torque, is also the one that is responsible for the spin pumping in dynamic RKKY theory. 
By taking into account the changes of $J_U$ and numerically evaluating Eq.~\ref{Eq.spinmixing}, we  show that the discrepancy of the increasing/decreasing values of spin mixing conductance in Ref.~\onlinecite{PhysRevLett.112.197201} and Ref.~\onlinecite{PhysRevB.94.014414} arises from the different strength of screening at the interface. 
In the case of nearly magnetic metals with strong screening of exchange interaction at the interface, the spin mixing conductance  is monotonically decreasing as the electron-electron interaction increase. Fig.~\ref{Fig.U} shows that a metallic ferromagnetic layer give strong screening while an insulating ferromagnetic layer give weak screening. Insulating ferromagnet enhances the  conductance for non-magnetic metal with small $UN(\epsilon_F)\ll 1$.

\section*{Acknowledgement}

We acknowledge financial support from Ministry of Research and Technology of the Republic of Indonesia through PDUPT Grant No. NKB-2816/UN2.RST/HKP.05.00/2020.

%\section*{DATA AVAILABiLITY STATEMENTS}
%Data sharing is not applicable to this article as no new data were created or analyzed in this study.

%
%\bibliography{ref}

\begin{thebibliography}{26}%
\makeatletter
\providecommand \@ifxundefined [1]{%
 \@ifx{#1\undefined}
}%
\providecommand \@ifnum [1]{%
 \ifnum #1\expandafter \@firstoftwo
 \else \expandafter \@secondoftwo
 \fi
}%
\providecommand \@ifx [1]{%
 \ifx #1\expandafter \@firstoftwo
 \else \expandafter \@secondoftwo
 \fi
}%
\providecommand \natexlab [1]{#1}%
\providecommand \enquote  [1]{``#1''}%
\providecommand \bibnamefont  [1]{#1}%
\providecommand \bibfnamefont [1]{#1}%
\providecommand \citenamefont [1]{#1}%
\providecommand \href@noop [0]{\@secondoftwo}%
\providecommand \href [0]{\begingroup \@sanitize@url \@href}%
\providecommand \@href[1]{\@@startlink{#1}\@@href}%
\providecommand \@@href[1]{\endgroup#1\@@endlink}%
\providecommand \@sanitize@url [0]{\catcode `\\12\catcode `\$12\catcode
  `\&12\catcode `\#12\catcode `\^12\catcode `\_12\catcode `\%12\relax}%
\providecommand \@@startlink[1]{}%
\providecommand \@@endlink[0]{}%
\providecommand \url  [0]{\begingroup\@sanitize@url \@url }%
\providecommand \@url [1]{\endgroup\@href {#1}{\urlprefix }}%
\providecommand \urlprefix  [0]{URL }%
\providecommand \Eprint [0]{\href }%
\providecommand \doibase [0]{http://dx.doi.org/}%
\providecommand \selectlanguage [0]{\@gobble}%
\providecommand \bibinfo  [0]{\@secondoftwo}%
\providecommand \bibfield  [0]{\@secondoftwo}%
\providecommand \translation [1]{[#1]}%
\providecommand \BibitemOpen [0]{}%
\providecommand \bibitemStop [0]{}%
\providecommand \bibitemNoStop [0]{.\EOS\space}%
\providecommand \EOS [0]{\spacefactor3000\relax}%
\providecommand \BibitemShut  [1]{\csname bibitem#1\endcsname}%
\let\auto@bib@innerbib\@empty
%</preamble>
\bibitem [{\citenamefont {Parkin}(1995)}]{reviewGMR}%
  \BibitemOpen
  \bibfield  {author} {\bibinfo {author} {\bibfnamefont {S.~S.~P.}\
  \bibnamefont {Parkin}},\ }\href {\doibase
  10.1146/annurev.ms.25.080195.002041} {\bibfield  {journal} {\bibinfo
  {journal} {Annual Review of Materials Science}\ }\textbf {\bibinfo {volume}
  {25}},\ \bibinfo {pages} {357} (\bibinfo {year} {1995})},\ \Eprint
  {http://arxiv.org/abs/https://doi.org/10.1146/annurev.ms.25.080195.002041}
  {https://doi.org/10.1146/annurev.ms.25.080195.002041} \BibitemShut {NoStop}%
\bibitem [{\citenamefont {Barna\ifmmode~\acute{s}\else \'{s}\fi{}}\ \emph
  {et~al.}(2005)\citenamefont {Barna\ifmmode~\acute{s}\else \'{s}\fi{}},
  \citenamefont {Fert}, \citenamefont {Gmitra}, \citenamefont {Weymann},\ and\
  \citenamefont {Dugaev}}]{PhysRevB.72.024426}%
  \BibitemOpen
  \bibfield  {author} {\bibinfo {author} {\bibfnamefont {J.}~\bibnamefont
  {Barna\ifmmode~\acute{s}\else \'{s}\fi{}}}, \bibinfo {author} {\bibfnamefont
  {A.}~\bibnamefont {Fert}}, \bibinfo {author} {\bibfnamefont {M.}~\bibnamefont
  {Gmitra}}, \bibinfo {author} {\bibfnamefont {I.}~\bibnamefont {Weymann}}, \
  and\ \bibinfo {author} {\bibfnamefont {V.~K.}\ \bibnamefont {Dugaev}},\
  }\href {\doibase 10.1103/PhysRevB.72.024426} {\bibfield  {journal} {\bibinfo
  {journal} {Phys. Rev. B}\ }\textbf {\bibinfo {volume} {72}},\ \bibinfo
  {pages} {024426} (\bibinfo {year} {2005})}\BibitemShut {NoStop}%
\bibitem [{\citenamefont {Brataas}\ \emph {et~al.}(2012)\citenamefont
  {Brataas}, \citenamefont {Tserkovnyak}, \citenamefont {Bauer},\ and\
  \citenamefont {Kelly}}]{SpinCurrent}%
  \BibitemOpen
  \bibfield  {author} {\bibinfo {author} {\bibfnamefont {A.}~\bibnamefont
  {Brataas}}, \bibinfo {author} {\bibfnamefont {Y.}~\bibnamefont
  {Tserkovnyak}}, \bibinfo {author} {\bibfnamefont {G.~E.~W.}\ \bibnamefont
  {Bauer}}, \ and\ \bibinfo {author} {\bibfnamefont {P.~J.}\ \bibnamefont
  {Kelly}},\ }\href@noop {} {\enquote {\bibinfo {title} {Spin pumping and spin
  transfer},}\ } (\bibinfo {year} {2012}),\ \Eprint
  {http://arxiv.org/abs/1108.0385} {arXiv:1108.0385 [cond-mat.mes-hall]}
  \BibitemShut {NoStop}%
\bibitem [{\citenamefont {Xiao}\ \emph {et~al.}(2008)\citenamefont {Xiao},
  \citenamefont {Bauer},\ and\ \citenamefont {Brataas}}]{PhysRevB.77.224419}%
  \BibitemOpen
  \bibfield  {author} {\bibinfo {author} {\bibfnamefont {J.}~\bibnamefont
  {Xiao}}, \bibinfo {author} {\bibfnamefont {G.~E.~W.}\ \bibnamefont {Bauer}},
  \ and\ \bibinfo {author} {\bibfnamefont {A.}~\bibnamefont {Brataas}},\ }\href
  {\doibase 10.1103/PhysRevB.77.224419} {\bibfield  {journal} {\bibinfo
  {journal} {Phys. Rev. B}\ }\textbf {\bibinfo {volume} {77}},\ \bibinfo
  {pages} {224419} (\bibinfo {year} {2008})}\BibitemShut {NoStop}%
\bibitem [{\citenamefont {Tserkovnyak}\ \emph
  {et~al.}(2002{\natexlab{a}})\citenamefont {Tserkovnyak}, \citenamefont
  {Brataas},\ and\ \citenamefont {Bauer}}]{PhysRevB.66.224403}%
  \BibitemOpen
  \bibfield  {author} {\bibinfo {author} {\bibfnamefont {Y.}~\bibnamefont
  {Tserkovnyak}}, \bibinfo {author} {\bibfnamefont {A.}~\bibnamefont
  {Brataas}}, \ and\ \bibinfo {author} {\bibfnamefont {G.~E.~W.}\ \bibnamefont
  {Bauer}},\ }\href {\doibase 10.1103/PhysRevB.66.224403} {\bibfield  {journal}
  {\bibinfo  {journal} {Phys. Rev. B}\ }\textbf {\bibinfo {volume} {66}},\
  \bibinfo {pages} {224403} (\bibinfo {year} {2002}{\natexlab{a}})}\BibitemShut
  {NoStop}%
\bibitem [{\citenamefont {\ifmmode~\check{S}\else
  \v{S}\fi{}im\'anek}(2003)}]{PhysRevB.68.224403}%
  \BibitemOpen
  \bibfield  {author} {\bibinfo {author} {\bibfnamefont {E.}~\bibnamefont
  {\ifmmode~\check{S}\else \v{S}\fi{}im\'anek}},\ }\href {\doibase
  10.1103/PhysRevB.68.224403} {\bibfield  {journal} {\bibinfo  {journal} {Phys.
  Rev. B}\ }\textbf {\bibinfo {volume} {68}},\ \bibinfo {pages} {224403}
  (\bibinfo {year} {2003})}\BibitemShut {NoStop}%
\bibitem [{\citenamefont {Cahaya}\ \emph {et~al.}(2017)\citenamefont {Cahaya},
  \citenamefont {Leon},\ and\ \citenamefont {Bauer}}]{PhysRevB.96.144434}%
  \BibitemOpen
  \bibfield  {author} {\bibinfo {author} {\bibfnamefont {A.~B.}\ \bibnamefont
  {Cahaya}}, \bibinfo {author} {\bibfnamefont {A.~O.}\ \bibnamefont {Leon}}, \
  and\ \bibinfo {author} {\bibfnamefont {G.~E.~W.}\ \bibnamefont {Bauer}},\
  }\href {\doibase 10.1103/PhysRevB.96.144434} {\bibfield  {journal} {\bibinfo
  {journal} {Phys. Rev. B}\ }\textbf {\bibinfo {volume} {96}},\ \bibinfo
  {pages} {144434} (\bibinfo {year} {2017})}\BibitemShut {NoStop}%
\bibitem [{\citenamefont {Tserkovnyak}\ \emph
  {et~al.}(2002{\natexlab{b}})\citenamefont {Tserkovnyak}, \citenamefont
  {Brataas},\ and\ \citenamefont {Bauer}}]{PhysRevLett.88.117601}%
  \BibitemOpen
  \bibfield  {author} {\bibinfo {author} {\bibfnamefont {Y.}~\bibnamefont
  {Tserkovnyak}}, \bibinfo {author} {\bibfnamefont {A.}~\bibnamefont
  {Brataas}}, \ and\ \bibinfo {author} {\bibfnamefont {G.~E.~W.}\ \bibnamefont
  {Bauer}},\ }\href {\doibase 10.1103/PhysRevLett.88.117601} {\bibfield
  {journal} {\bibinfo  {journal} {Phys. Rev. Lett.}\ }\textbf {\bibinfo
  {volume} {88}},\ \bibinfo {pages} {117601} (\bibinfo {year}
  {2002}{\natexlab{b}})}\BibitemShut {NoStop}%
\bibitem [{\citenamefont {Carva}\ and\ \citenamefont
  {Turek}(2007)}]{PhysRevB.76.104409}%
  \BibitemOpen
  \bibfield  {author} {\bibinfo {author} {\bibfnamefont {K.}~\bibnamefont
  {Carva}}\ and\ \bibinfo {author} {\bibfnamefont {I.}~\bibnamefont {Turek}},\
  }\href {\doibase 10.1103/PhysRevB.76.104409} {\bibfield  {journal} {\bibinfo
  {journal} {Phys. Rev. B}\ }\textbf {\bibinfo {volume} {76}},\ \bibinfo
  {pages} {104409} (\bibinfo {year} {2007})}\BibitemShut {NoStop}%
\bibitem [{\citenamefont {Weiler}\ \emph {et~al.}(2013)\citenamefont {Weiler},
  \citenamefont {Althammer}, \citenamefont {Schreier}, \citenamefont {Lotze},
  \citenamefont {Pernpeintner}, \citenamefont {Meyer}, \citenamefont {Huebl},
  \citenamefont {Gross}, \citenamefont {Kamra}, \citenamefont {Xiao},
  \citenamefont {Chen}, \citenamefont {Jiao}, \citenamefont {Bauer},\ and\
  \citenamefont {Goennenwein}}]{PhysRevLett.111.176601}%
  \BibitemOpen
  \bibfield  {author} {\bibinfo {author} {\bibfnamefont {M.}~\bibnamefont
  {Weiler}}, \bibinfo {author} {\bibfnamefont {M.}~\bibnamefont {Althammer}},
  \bibinfo {author} {\bibfnamefont {M.}~\bibnamefont {Schreier}}, \bibinfo
  {author} {\bibfnamefont {J.}~\bibnamefont {Lotze}}, \bibinfo {author}
  {\bibfnamefont {M.}~\bibnamefont {Pernpeintner}}, \bibinfo {author}
  {\bibfnamefont {S.}~\bibnamefont {Meyer}}, \bibinfo {author} {\bibfnamefont
  {H.}~\bibnamefont {Huebl}}, \bibinfo {author} {\bibfnamefont
  {R.}~\bibnamefont {Gross}}, \bibinfo {author} {\bibfnamefont
  {A.}~\bibnamefont {Kamra}}, \bibinfo {author} {\bibfnamefont
  {J.}~\bibnamefont {Xiao}}, \bibinfo {author} {\bibfnamefont {Y.-T.}\
  \bibnamefont {Chen}}, \bibinfo {author} {\bibfnamefont {H.~J.}\ \bibnamefont
  {Jiao}}, \bibinfo {author} {\bibfnamefont {G.~E.~W.}\ \bibnamefont {Bauer}},
  \ and\ \bibinfo {author} {\bibfnamefont {S.~T.~B.}\ \bibnamefont
  {Goennenwein}},\ }\href {\doibase 10.1103/PhysRevLett.111.176601} {\bibfield
  {journal} {\bibinfo  {journal} {Phys. Rev. Lett.}\ }\textbf {\bibinfo
  {volume} {111}},\ \bibinfo {pages} {176601} (\bibinfo {year}
  {2013})}\BibitemShut {NoStop}%
\bibitem [{\citenamefont {Sigalas}\ and\ \citenamefont
  {Papaconstantopoulos}(1994)}]{PhysRevB.50.7255}%
  \BibitemOpen
  \bibfield  {author} {\bibinfo {author} {\bibfnamefont {M.~M.}\ \bibnamefont
  {Sigalas}}\ and\ \bibinfo {author} {\bibfnamefont {D.~A.}\ \bibnamefont
  {Papaconstantopoulos}},\ }\href {\doibase 10.1103/PhysRevB.50.7255}
  {\bibfield  {journal} {\bibinfo  {journal} {Phys. Rev. B}\ }\textbf {\bibinfo
  {volume} {50}},\ \bibinfo {pages} {7255} (\bibinfo {year}
  {1994})}\BibitemShut {NoStop}%
\bibitem [{\citenamefont {Zellermann}\ \emph {et~al.}(2004)\citenamefont
  {Zellermann}, \citenamefont {Paintner},\ and\ \citenamefont
  {Voitländer}}]{Zellermann_2004}%
  \BibitemOpen
  \bibfield  {author} {\bibinfo {author} {\bibfnamefont {B.}~\bibnamefont
  {Zellermann}}, \bibinfo {author} {\bibfnamefont {A.}~\bibnamefont
  {Paintner}}, \ and\ \bibinfo {author} {\bibfnamefont {J.}~\bibnamefont
  {Voitländer}},\ }\href {\doibase 10.1088/0953-8984/16/6/019} {\bibfield
  {journal} {\bibinfo  {journal} {Journal of Physics: Condensed Matter}\
  }\textbf {\bibinfo {volume} {16}},\ \bibinfo {pages} {919} (\bibinfo {year}
  {2004})}\BibitemShut {NoStop}%
\bibitem [{\citenamefont {Povzner}\ \emph {et~al.}(2010)\citenamefont
  {Povzner}, \citenamefont {Volkov},\ and\ \citenamefont
  {Filanovich}}]{Povzner2010}%
  \BibitemOpen
  \bibfield  {author} {\bibinfo {author} {\bibfnamefont {A.~A.}\ \bibnamefont
  {Povzner}}, \bibinfo {author} {\bibfnamefont {A.~G.}\ \bibnamefont {Volkov}},
  \ and\ \bibinfo {author} {\bibfnamefont {A.~N.}\ \bibnamefont {Filanovich}},\
  }\href {\doibase 10.1134/S1063783410100021} {\bibfield  {journal} {\bibinfo
  {journal} {Physics of the Solid State}\ }\textbf {\bibinfo {volume} {52}},\
  \bibinfo {pages} {2012} (\bibinfo {year} {2010})}\BibitemShut {NoStop}%
\bibitem [{\citenamefont {Santos}\ \emph {et~al.}(2013)\citenamefont {Santos},
  \citenamefont {Venezuela}, \citenamefont {Muniz},\ and\ \citenamefont
  {Costa}}]{PhysRevB.88.054423}%
  \BibitemOpen
  \bibfield  {author} {\bibinfo {author} {\bibfnamefont {D.~L.~R.}\
  \bibnamefont {Santos}}, \bibinfo {author} {\bibfnamefont {P.}~\bibnamefont
  {Venezuela}}, \bibinfo {author} {\bibfnamefont {R.~B.}\ \bibnamefont
  {Muniz}}, \ and\ \bibinfo {author} {\bibfnamefont {A.~T.}\ \bibnamefont
  {Costa}},\ }\href {\doibase 10.1103/PhysRevB.88.054423} {\bibfield  {journal}
  {\bibinfo  {journal} {Phys. Rev. B}\ }\textbf {\bibinfo {volume} {88}},\
  \bibinfo {pages} {054423} (\bibinfo {year} {2013})}\BibitemShut {NoStop}%
\bibitem [{\citenamefont {\ifmmode~\check{S}\else \v{S}\fi{}im\'anek}\ and\
  \citenamefont {Heinrich}(2003)}]{PhysRevB.67.144418}%
  \BibitemOpen
  \bibfield  {author} {\bibinfo {author} {\bibfnamefont {E.}~\bibnamefont
  {\ifmmode~\check{S}\else \v{S}\fi{}im\'anek}}\ and\ \bibinfo {author}
  {\bibfnamefont {B.}~\bibnamefont {Heinrich}},\ }\href {\doibase
  10.1103/PhysRevB.67.144418} {\bibfield  {journal} {\bibinfo  {journal} {Phys.
  Rev. B}\ }\textbf {\bibinfo {volume} {67}},\ \bibinfo {pages} {144418}
  (\bibinfo {year} {2003})}\BibitemShut {NoStop}%
\bibitem [{\citenamefont {Wang}\ \emph {et~al.}(2014)\citenamefont {Wang},
  \citenamefont {Du}, \citenamefont {Pu}, \citenamefont {Adur}, \citenamefont
  {Hammel},\ and\ \citenamefont {Yang}}]{PhysRevLett.112.197201}%
  \BibitemOpen
  \bibfield  {author} {\bibinfo {author} {\bibfnamefont {H.~L.}\ \bibnamefont
  {Wang}}, \bibinfo {author} {\bibfnamefont {C.~H.}\ \bibnamefont {Du}},
  \bibinfo {author} {\bibfnamefont {Y.}~\bibnamefont {Pu}}, \bibinfo {author}
  {\bibfnamefont {R.}~\bibnamefont {Adur}}, \bibinfo {author} {\bibfnamefont
  {P.~C.}\ \bibnamefont {Hammel}}, \ and\ \bibinfo {author} {\bibfnamefont
  {F.~Y.}\ \bibnamefont {Yang}},\ }\href {\doibase
  10.1103/PhysRevLett.112.197201} {\bibfield  {journal} {\bibinfo  {journal}
  {Phys. Rev. Lett.}\ }\textbf {\bibinfo {volume} {112}},\ \bibinfo {pages}
  {197201} (\bibinfo {year} {2014})}\BibitemShut {NoStop}%
\bibitem [{\citenamefont {Caminale}\ \emph {et~al.}(2016)\citenamefont
  {Caminale}, \citenamefont {Ghosh}, \citenamefont {Auffret}, \citenamefont
  {Ebels}, \citenamefont {Ollefs}, \citenamefont {Wilhelm}, \citenamefont
  {Rogalev},\ and\ \citenamefont {Bailey}}]{PhysRevB.94.014414}%
  \BibitemOpen
  \bibfield  {author} {\bibinfo {author} {\bibfnamefont {M.}~\bibnamefont
  {Caminale}}, \bibinfo {author} {\bibfnamefont {A.}~\bibnamefont {Ghosh}},
  \bibinfo {author} {\bibfnamefont {S.}~\bibnamefont {Auffret}}, \bibinfo
  {author} {\bibfnamefont {U.}~\bibnamefont {Ebels}}, \bibinfo {author}
  {\bibfnamefont {K.}~\bibnamefont {Ollefs}}, \bibinfo {author} {\bibfnamefont
  {F.}~\bibnamefont {Wilhelm}}, \bibinfo {author} {\bibfnamefont
  {A.}~\bibnamefont {Rogalev}}, \ and\ \bibinfo {author} {\bibfnamefont
  {W.~E.}\ \bibnamefont {Bailey}},\ }\href {\doibase
  10.1103/PhysRevB.94.014414} {\bibfield  {journal} {\bibinfo  {journal} {Phys.
  Rev. B}\ }\textbf {\bibinfo {volume} {94}},\ \bibinfo {pages} {014414}
  (\bibinfo {year} {2016})}\BibitemShut {NoStop}%
\bibitem [{\citenamefont {Xiao}\ \emph {et~al.}(2010)\citenamefont {Xiao},
  \citenamefont {Bauer}, \citenamefont {Uchida}, \citenamefont {Saitoh},\ and\
  \citenamefont {Maekawa}}]{PhysRevB.81.214418}%
  \BibitemOpen
  \bibfield  {author} {\bibinfo {author} {\bibfnamefont {J.}~\bibnamefont
  {Xiao}}, \bibinfo {author} {\bibfnamefont {G.~E.~W.}\ \bibnamefont {Bauer}},
  \bibinfo {author} {\bibfnamefont {K.-c.}\ \bibnamefont {Uchida}}, \bibinfo
  {author} {\bibfnamefont {E.}~\bibnamefont {Saitoh}}, \ and\ \bibinfo {author}
  {\bibfnamefont {S.}~\bibnamefont {Maekawa}},\ }\href {\doibase
  10.1103/PhysRevB.81.214418} {\bibfield  {journal} {\bibinfo  {journal} {Phys.
  Rev. B}\ }\textbf {\bibinfo {volume} {81}},\ \bibinfo {pages} {214418}
  (\bibinfo {year} {2010})}\BibitemShut {NoStop}%
\bibitem [{\citenamefont {{Cahaya}}\ \emph {et~al.}(2015)\citenamefont
  {{Cahaya}}, \citenamefont {{Tretiakov}},\ and\ \citenamefont
  {{Bauer}}}]{7111309}%
  \BibitemOpen
  \bibfield  {author} {\bibinfo {author} {\bibfnamefont {A.~B.}\ \bibnamefont
  {{Cahaya}}}, \bibinfo {author} {\bibfnamefont {O.~A.}\ \bibnamefont
  {{Tretiakov}}}, \ and\ \bibinfo {author} {\bibfnamefont {G.~E.~W.}\
  \bibnamefont {{Bauer}}},\ }\href@noop {} {\bibfield  {journal} {\bibinfo
  {journal} {IEEE Transactions on Magnetics}\ }\textbf {\bibinfo {volume}
  {51}},\ \bibinfo {pages} {1} (\bibinfo {year} {2015})}\BibitemShut {NoStop}%
\bibitem [{\citenamefont {Spiesser}\ \emph {et~al.}(2017)\citenamefont
  {Spiesser}, \citenamefont {Saito}, \citenamefont {Fujita}, \citenamefont
  {Yamada}, \citenamefont {Hamaya}, \citenamefont {Yuasa},\ and\ \citenamefont
  {Jansen}}]{PhysRevApplied.8.064023}%
  \BibitemOpen
  \bibfield  {author} {\bibinfo {author} {\bibfnamefont {A.}~\bibnamefont
  {Spiesser}}, \bibinfo {author} {\bibfnamefont {H.}~\bibnamefont {Saito}},
  \bibinfo {author} {\bibfnamefont {Y.}~\bibnamefont {Fujita}}, \bibinfo
  {author} {\bibfnamefont {S.}~\bibnamefont {Yamada}}, \bibinfo {author}
  {\bibfnamefont {K.}~\bibnamefont {Hamaya}}, \bibinfo {author} {\bibfnamefont
  {S.}~\bibnamefont {Yuasa}}, \ and\ \bibinfo {author} {\bibfnamefont
  {R.}~\bibnamefont {Jansen}},\ }\href {\doibase
  10.1103/PhysRevApplied.8.064023} {\bibfield  {journal} {\bibinfo  {journal}
  {Phys. Rev. Applied}\ }\textbf {\bibinfo {volume} {8}},\ \bibinfo {pages}
  {064023} (\bibinfo {year} {2017})}\BibitemShut {NoStop}%
\bibitem [{\citenamefont {Takahashi}\ and\ \citenamefont
  {Maekawa}(2008)}]{doi:10.1088/1468-6996/9/1/014105}%
  \BibitemOpen
  \bibfield  {author} {\bibinfo {author} {\bibfnamefont {S.}~\bibnamefont
  {Takahashi}}\ and\ \bibinfo {author} {\bibfnamefont {S.}~\bibnamefont
  {Maekawa}},\ }\href {\doibase 10.1088/1468-6996/9/1/014105} {\bibfield
  {journal} {\bibinfo  {journal} {Science and Technology of Advanced
  Materials}\ }\textbf {\bibinfo {volume} {9}},\ \bibinfo {pages} {014105}
  (\bibinfo {year} {2008})},\ \bibinfo {note} {pMID: 27877931}\BibitemShut
  {NoStop}%
\bibitem [{\citenamefont {Kim}(1999)}]{BookKim}%
  \BibitemOpen
  \bibfield  {author} {\bibinfo {author} {\bibfnamefont {D.~J.}\ \bibnamefont
  {Kim}},\ }\href@noop {} {\emph {\bibinfo {title} {New Perspectives in
  Magnetism of Metals}}}\ (\bibinfo  {publisher} {Springer Science+Business
  Media},\ \bibinfo {address} {New York},\ \bibinfo {year} {1999})\BibitemShut
  {NoStop}%
\bibitem [{\citenamefont {Cahaya}\ \emph {et~al.}(2020)\citenamefont {Cahaya},
  \citenamefont {Azhar},\ and\ \citenamefont {Majidi}}]{Racah}%
  \BibitemOpen
  \bibfield  {author} {\bibinfo {author} {\bibfnamefont {A.~B.}\ \bibnamefont
  {Cahaya}}, \bibinfo {author} {\bibfnamefont {A.}~\bibnamefont {Azhar}}, \
  and\ \bibinfo {author} {\bibfnamefont {M.~A.}\ \bibnamefont {Majidi}},\
  }\href {\doibase https://doi.org/10.1016/j.physb.2020.412696} {\bibfield
  {journal} {\bibinfo  {journal} {Physica B: Condensed Matter}\ ,\ \bibinfo
  {pages} {412696}} (\bibinfo {year} {2020})}\BibitemShut {NoStop}%
\bibitem [{\citenamefont {Jiao}\ and\ \citenamefont {Ho}(2015)}]{JIAO2015140}%
  \BibitemOpen
  \bibfield  {author} {\bibinfo {author} {\bibfnamefont {L.~G.}\ \bibnamefont
  {Jiao}}\ and\ \bibinfo {author} {\bibfnamefont {Y.~K.}\ \bibnamefont {Ho}},\
  }\href {\doibase https://doi.org/10.1016/j.cpc.2014.11.019} {\bibfield
  {journal} {\bibinfo  {journal} {Computer Physics Communications}\ }\textbf
  {\bibinfo {volume} {188}},\ \bibinfo {pages} {140 } (\bibinfo {year}
  {2015})}\BibitemShut {NoStop}%
\bibitem [{\citenamefont {Ba{\u{g}}c{\i}}\ \emph {et~al.}(2018)\citenamefont
  {Ba{\u{g}}c{\i}}, \citenamefont {Hoggan},\ and\ \citenamefont
  {Adak}}]{Bagci2018}%
  \BibitemOpen
  \bibfield  {author} {\bibinfo {author} {\bibfnamefont {A.}~\bibnamefont
  {Ba{\u{g}}c{\i}}}, \bibinfo {author} {\bibfnamefont {P.~E.}\ \bibnamefont
  {Hoggan}}, \ and\ \bibinfo {author} {\bibfnamefont {M.}~\bibnamefont
  {Adak}},\ }\href {\doibase 10.1007/s12210-018-0734-3} {\bibfield  {journal}
  {\bibinfo  {journal} {Rendiconti Lincei. Scienze Fisiche e Naturali}\
  }\textbf {\bibinfo {volume} {29}},\ \bibinfo {pages} {765} (\bibinfo {year}
  {2018})}\BibitemShut {NoStop}%
\bibitem [{\citenamefont {Ribic}\ \emph {et~al.}(2014)\citenamefont {Ribic},
  \citenamefont {Assmann}, \citenamefont {T\'oth},\ and\ \citenamefont
  {Held}}]{PhysRevB.90.165105}%
  \BibitemOpen
  \bibfield  {author} {\bibinfo {author} {\bibfnamefont {T.}~\bibnamefont
  {Ribic}}, \bibinfo {author} {\bibfnamefont {E.}~\bibnamefont {Assmann}},
  \bibinfo {author} {\bibfnamefont {A.}~\bibnamefont {T\'oth}}, \ and\ \bibinfo
  {author} {\bibfnamefont {K.}~\bibnamefont {Held}},\ }\href {\doibase
  10.1103/PhysRevB.90.165105} {\bibfield  {journal} {\bibinfo  {journal} {Phys.
  Rev. B}\ }\textbf {\bibinfo {volume} {90}},\ \bibinfo {pages} {165105}
  (\bibinfo {year} {2014})}\BibitemShut {NoStop}%
\end{thebibliography}

\end{document}